\documentclass{llncs}
\usepackage{mathptmx}       
\usepackage{helvet}         
\usepackage{courier}        
\usepackage{type1cm}        
\usepackage{makeidx}         
\usepackage{graphicx}        
\usepackage{multicol}        
\usepackage[bottom]{footmisc}
\usepackage{amsfonts}
\usepackage{amssymb}
\usepackage[cmex10]{amsmath}
\usepackage{algorithm}
\usepackage{algpseudocode}
\usepackage{subfig}

\usepackage{graphicx}
\usepackage{booktabs}
\usepackage{amssymb}
\usepackage{amsmath}
 \usepackage{MnSymbol}
 \usepackage{mathtools}
 \usepackage{upgreek}
\usepackage{lipsum} 
\usepackage{tabularx}
\begin{document}

\title{A logical approach for temporal and multiplex networks analysis \vspace{-12pt}} 

\author{Esteban Bautista and Matthieu Latapy}
\authorrunning{Ivar Ekeland et al.} 
%
%
\institute{Sorbonne Universit\'{e}, CNRS, LIP6, F-75005 Paris, France \vspace{-15pt}}
\maketitle    

\section{Introduction}  \vspace{-10pt}
Many systems generate data as a set of triplets $(a, b, c)$: they may represent that user $a$ called $b$ at time $c$ or that customer $a$ purchased product $b$ in store $c$. These datasets are traditionally studied as networks with an extra dimension (time or layer), for which the fields of temporal and multiplex networks have extended graph theory to account for the new dimension \cite{Kivela}. However, such frameworks detach one variable from the others and allow to extend one same concept in many ways, making it hard to capture patterns across all dimensions and to identify the best definitions for a given dataset. This work overrides this vision and proposes a direct processing of the set of triplets. While \cite{Cerf} also approaches triplets directly, it focuses on specific patterns and applications. Our work shows that a more general analysis is possible by partitioning the data and building categorical propositions (CPs) that encode informative patterns. We show that several concepts from graph theory can be framed under this formalism and leverage such insights to extend the concepts to data triplets. Lastly, we propose an algorithm to list CPs satisfying specific constraints and apply it to a real world dataset. \vspace{-10pt}

\section{Results}
\begin{figure}[t!]
\includegraphics[width=0.5\linewidth]{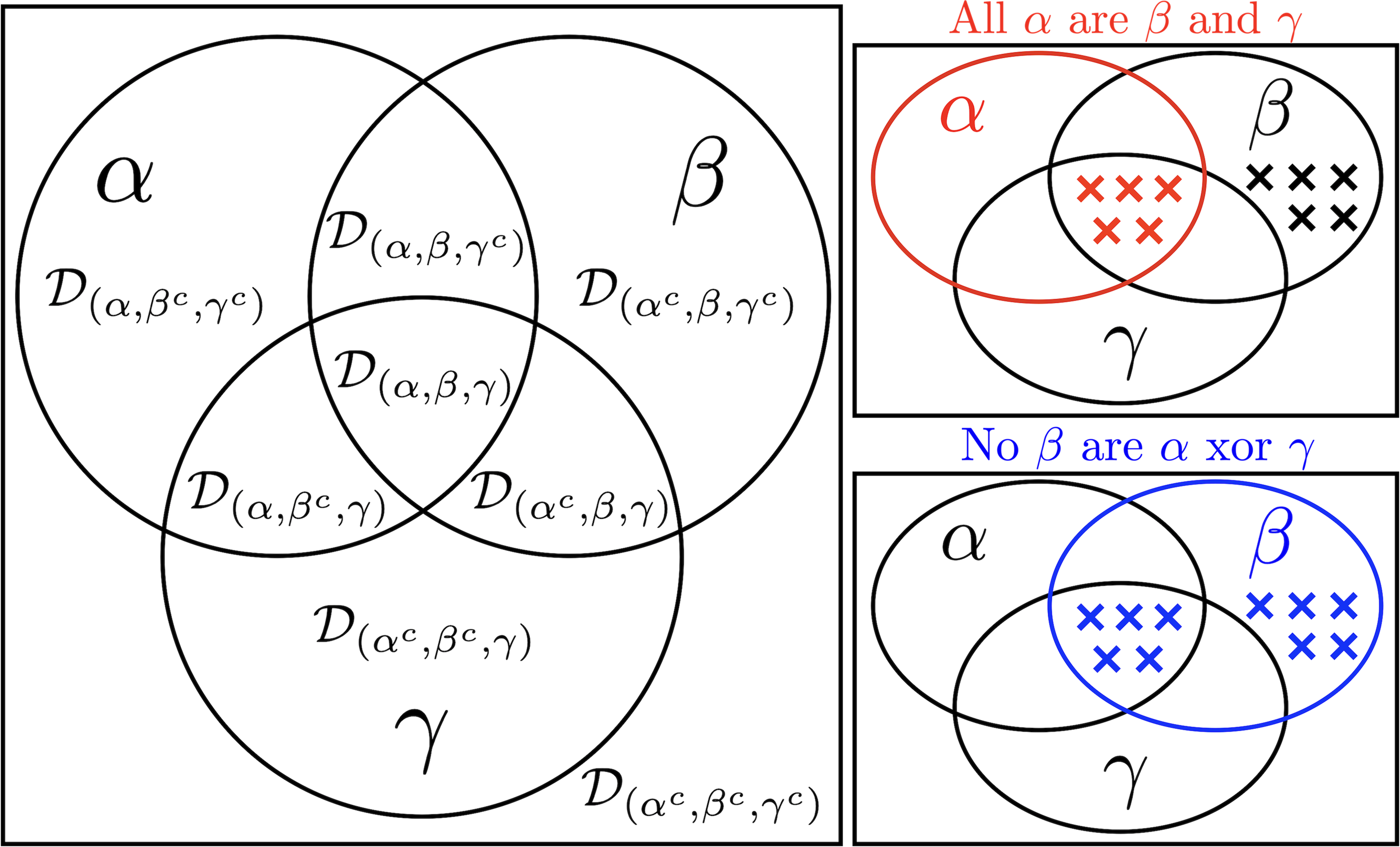} \hspace{50pt}
\includegraphics[width=0.28\linewidth]{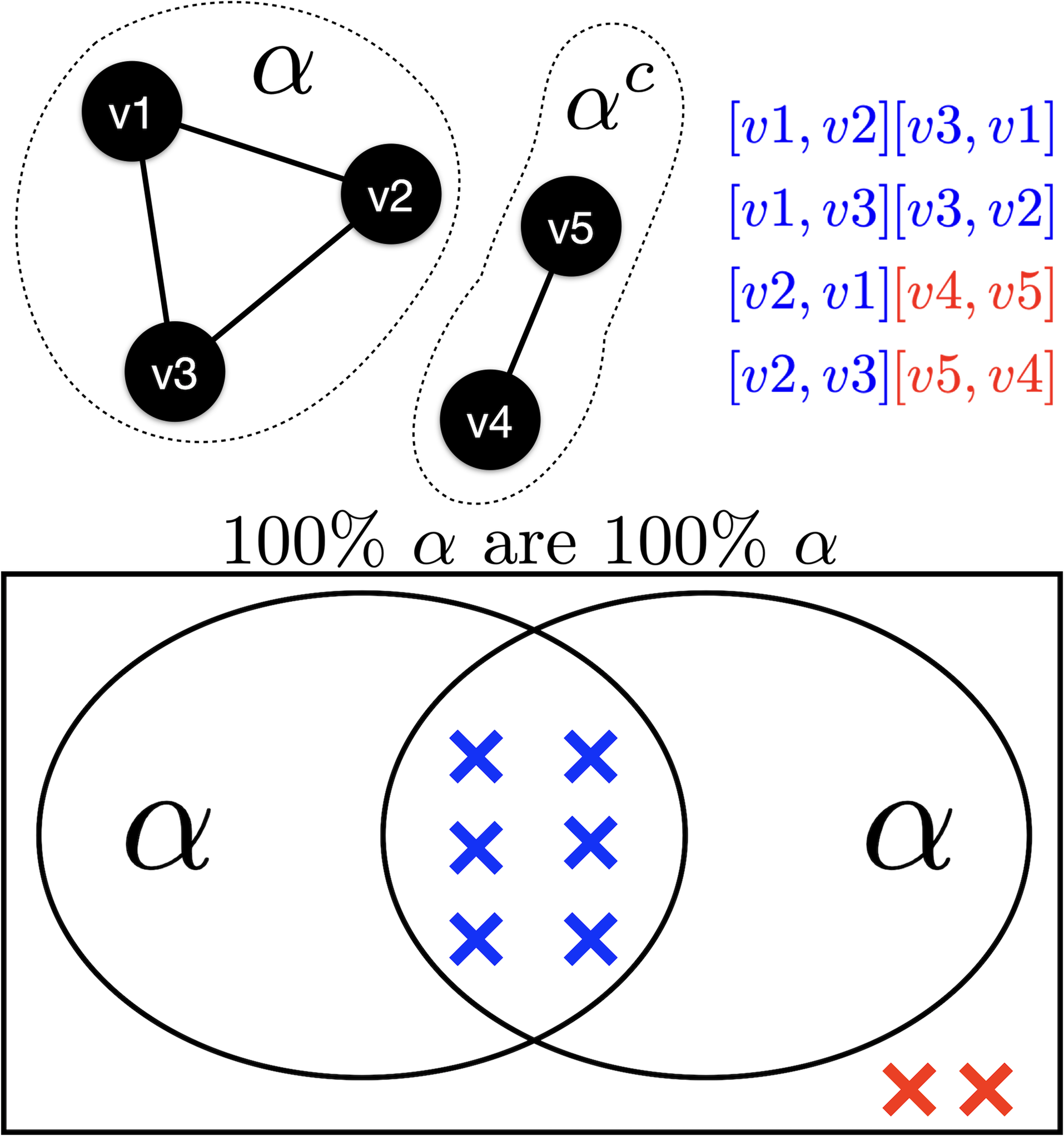} 
\caption{ {\bf Left}: Partition bins (big square) and distribution of $\mathcal{D}$ (crosses) into the bins with associated categorical propositions (small squares). {\bf Right}: Framework applied to a graph composed of two clique components.} \vspace{-20pt}
\label{fig_prop}
\end{figure}

\vspace{-10pt}
{\bf Analysis via propositions}. We consider the most general case where all the triplet entries come from arbitrary sets $A, B, C$. We thus define a triplet space as $\mathcal{S} = \{(a,b,c) | a \in A, b \in B, c \in C \}$ and a dataset as $\mathcal{D} \subseteq \mathcal{S}$. We also define the sub-dataset induced by $\alpha \subseteq A$, $\beta \subseteq B$, $\gamma \subseteq C$ as $\mathcal{D}_{(\alpha, \beta, \gamma)} = \{(a,b,c) \in \mathcal{D} | a \in \alpha, b \in \beta, c \in \gamma \}$. Our main observation is that given $\alpha \subseteq A$, $\beta \subseteq B$, $\gamma \subseteq C$, we can partition $\mathcal{D}$ into eight disjoint regions (or bins) according to whether a triplet has its entries in $\alpha$, $\beta$ and $\gamma$. Then, we can capture how triplets in $\mathcal{D}$ distribute across these bins via CPs. This process is illustrated in the Fig \ref{fig_prop}-Left: the large square depicts the eight possible partition bins, while the smaller squares illustrate how the triplets (crosses) may distribute and CPs be constructed to capture the distribution pattern. In a nutshell, a CP asserts or denies that all or some of the members of one group (the subject) possess the attributes of another group (the predicate), using an expression of the form: `Q S are P', where S refers to the subject, P to the predicate, and Q to a quantification word which can be `All', `Some', or `No' \cite{Copi}. The expression `All S are P' is a typical example. In our case, we form CPs using $\alpha$, $\beta$, or $\gamma$ as S and the other two as P, such that the following expression holds: `Q (triplets with elements in) S are (in relation with at least one element from) P'. For simplicity, we omit the words in parenthesis. In Fig. \ref{fig_prop}-Left we notice that all the triplets with elements in $\alpha$ also have elements in $\beta$ and $\gamma$, thus forming: `All $\alpha$ are $\beta$ and $\gamma$'. These are informative patterns: if $\alpha$ represents customers, $\beta$ products and $\gamma$ stores, then `All $\alpha$ are $\beta$ and $\gamma$' indicates that customers in $\alpha$ buy only products from $\beta$ and only in stores from $\gamma$. It is thus of interest to list informative CPs. We notice that (i) universal quantifiers (All, No) are more informative than particular ones (Some), yet particular propositions may be close to a universal one; and (ii) propositions above do not express how dense are the relations between S and P. We therefore extend propositions to: $x \%$ S are $y \%$ P, where $x$ is the fraction of triplets in S in relation to P and $y$ is the density of relationships between S and P. This allows us to state the algorithmic challenge of listing all propositions satisfying constraints on $x$ and $y$ without needing to explore the full space. 
\newline
{\bf Relation to graph theory and extensions}. Our formalism can also be used to study tuples $(a, b)$. By setting $A = B = \mathcal{V}$ we address the particular case of graphs, where $\mathcal{V}$ is to the vertex set of graph $\mathcal{G}$. Several concepts from graph theory may be re-formulated in terms of propositions satisfying specific constraints. An illustration is given in Fig. \ref{fig_prop}-Right. Going further, we use this re-formulation to generalize the concepts to data triplets. Our results are listed in Table \ref{table_prop}. It can be seen that some patterns, like XOR predicates, may not be easily derived from pure graph extensions. 
\newline
{\bf Listing propositions}. We propose Algorithm \ref{algo_prop} to list propositions of type $x \geq x_{min} \%$ $\alpha$ are $y \geq y_{min} \%$ $\beta$ and $\gamma$, where $x_{min}$ and $y_{min}$ are user-defined parameters. It uses the fact that disjoint subjects satisfying a predicate also satisfy it when merged. Thus, the algorithm searches valid predicates for singleton subjects and merges all those sharing predicates. We find predicates via a constructive approach where each triplet forms a region iteratively grown until the constraints are no longer satisfied. While this approach does not in general retrieve all propositions, it identifies a significant number of non-trivial patterns, and it may be improved in further work.
\newline
{\bf Application to real-world data}. We apply Algorithm \ref{algo_prop} to a contact network in a hospital \cite{Vanhems}. Sets A = B consist of 29 patients and 46 healthcare workers (27 nurses, 11 doctors, 8 admin), set C represents time (1890 minutes of data). We use time as subject set and $x_{min} = 0.7, y_{min} = 0.5$. The algorithm finds 1456 predicates from which it forms patterns like: (i) group of 7 minutes where 85\% activity corresponds to 3 nurses and 1 admin interacting with 64\% density; (ii) group of 3 minutes where 84\% activity corresponds to 4 doctors interacting with 66\% density; (iii) group of 7 minutes where 80\% activity corresponds to 3 nurses interacting with 66\% density. Clearly, the patterns found are representative of the typical activity in a hospital.
\newline 
{\bf Acknowledgements}. This work is funded in part by the ANR (French National Agency of Research) under the Limass (ANR-19-CE23-0010) and FiT LabCom grants. 

\begin{table}[h]
\centering
\scriptsize
\begin{tabular}{ >{\raggedright\arraybackslash}m{15mm} m{35mm} m{65mm} }
\multicolumn{1}{ >{\centering\arraybackslash}m{15mm} }{\textbf{Concept}} 
    & \multicolumn{1}{>{\centering\arraybackslash}m{35mm}}{\textbf{Proposition in graphs}} 
    & \multicolumn{1}{>{\centering\arraybackslash}m{65mm}}{\textbf{Extension}} \\ \toprule
Disconnected network &
$\mathcal{G}$ is disconnected if there exists $\alpha \subset \mathcal{V}$ satisfying: \vspace{3pt} \newline 
\begin{tabular}{>{\centering\arraybackslash}m{35mm}}
       No $\alpha$ are $\alpha^c$ 
\end{tabular} &
$\mathcal{D}$ is disconnected if there exist $\alpha \subset A$, $\beta \subset B$, $\gamma \subset C$ satisfying: \vspace{3pt} \newline
\begin{tabular}{>{\centering\arraybackslash}m{32.5mm}m{32.5mm} }
         No $\alpha$ are $\beta^c$ or $\gamma^c$ & No $\alpha^c$ are $\beta$ or $\gamma$ \\
	No $\beta$ are $\alpha^c$ or $\beta^c$ &  No $\beta^c$ are $\alpha$ or $\gamma$ \\ 
	No $\gamma$ are $\alpha^c$ or $\beta^c$ & No $\gamma^c$ are  $\alpha$ or $\beta$ \\
\end{tabular}  \\ \midrule

Vertex cover  & 
$\alpha \subseteq \mathcal{V}$ is a cover of $\mathcal{G}$ if it is satisfied: \vspace{3pt} \newline
\begin{tabular}{>{\centering\arraybackslash}m{35mm}}
       No $\alpha^c$ are $\alpha^c$
\end{tabular} &
$\alpha \subseteq A$, $\beta \subseteq B$, and $\gamma \subseteq C$ are a cover of $\mathcal{D}$ if it is satisfied: \vspace{3pt} \newline
\begin{tabular}{>{\centering\arraybackslash}m{32.5mm}m{32.5mm} }
        No $\alpha^c$ are $\beta^c$ and $\gamma^c$ & No $\beta^c$ are $\alpha^c$ and $\beta^c$  \\  
	No $\gamma^c$ are $\alpha^c$ and $\beta^c$
\end{tabular}  \\ \midrule

Dominating set &
$\alpha \subseteq \mathcal{V}$ is a dominating set of $\mathcal{G}$ if it is satisfied: \vspace{3pt} \newline
\begin{tabular}{>{\centering\arraybackslash}m{35mm}}
Some $a_i$ are $\alpha$, $\forall a_i \in \alpha^c$
\end{tabular} & 
$\alpha \subseteq A$, $\beta \subseteq B$, and $\gamma \subseteq C$ are dominating sets of $\mathcal{D}$ if it is satisfied: \vspace{3pt} \newline
\begin{tabular}{>{\centering\arraybackslash}m{32.5mm}m{32.5mm} }
Some $a_i$ are $\beta$ or $\gamma$, $\forall a_i \in \alpha^c$  & Some $b_i$ are $\alpha$ or $\gamma$, $\forall b_i \in \beta^c$  \\ 
Some $c_i$ are $\alpha$ or $\beta$, $\forall c_i \in \gamma^c$ 
\end{tabular}  \\ \midrule

Separating set &
$\alpha \subseteq \mathcal{V}$ is a separating set of $\mathcal{G}$ if it is satisfied that:  \vspace{3pt} \newline
\begin{tabular}{>{\centering\arraybackslash}m{35mm}}
$\mathcal{G}_{(\alpha^c)}$ is disconnected 
\end{tabular} & 
$\alpha \subseteq A$, $\beta \subseteq B$, and $\gamma \subseteq C$ are separating sets of $\mathcal{D}$ if it is satisfied that:  \vspace{3pt} \newline
\begin{tabular}{>{\centering\arraybackslash}m{65mm} }
$\mathcal{D}_{(\alpha^c, \beta^c, \gamma^c)}$ is disconnected 
\end{tabular} \\ \midrule

Vertex k-coloring &
A disjoint partitioning $\mathcal{V} = \bigcup_{i = 1}^{k} \alpha_i$ is a k-coloring of $\mathcal{G}$ if it is satisfied: \vspace{3pt} \newline
\begin{tabular}{>{\centering\arraybackslash}m{35mm}}
No $\alpha_i$ are $\alpha_i$, $\forall i$
\end{tabular} &
The disjoint partitionings $A = \bigcup_{i = 1}^{k} \alpha_i$, $B = \bigcup_{i = 1}^{k} \beta_i$, $C = \bigcup_{i = 1}^{k} \gamma_i$ are a k-coloring of $\mathcal{D}$ if it is satisfied: \vspace{3pt} \newline
\begin{tabular}{>{\centering\arraybackslash}m{32.5mm}m{32.5mm} }
No $\alpha_i$ are $\beta_i$ and $\gamma_i$, $\forall i$ & No $\beta_i$ are $\alpha_i$ and $\gamma_i$, $\forall i$ \\
No $\gamma_i$ are $\alpha_i$ and $\beta_i$, $\forall i$
\end{tabular} \\ \midrule

Clique & 
$\alpha \subseteq \mathcal{V}$ is a clique of $\mathcal{G}$ if it satisfies: \vspace{3pt} \newline
\begin{tabular}{>{\centering\arraybackslash}m{35mm}}
Some $\alpha$ are $100\%$ $\alpha$
\end{tabular} &
$\alpha \subseteq A$, $\beta \subseteq B$, and $\gamma \subseteq C$ are a clique of $\mathcal{D}$ if they satisfy: \vspace{3pt} \newline
\begin{tabular}{>{\centering\arraybackslash}m{65mm}}
Some $\alpha$ are $100\%$ $\beta$ and $\gamma$
\end{tabular} \\ \midrule

Cluster & 
$\alpha \subseteq \mathcal{V}$ is a cluster of $\mathcal{G}$ if it satisfies: \vspace{3pt} \newline
\begin{tabular}{>{\centering\arraybackslash}m{40mm}}
Large x \% $\alpha$ are $\alpha$
\end{tabular} &
$\alpha \subseteq A$, $\beta \subseteq B$, and $\gamma \subseteq C$ are a cluster of $\mathcal{D}$ if they satisfy: \vspace{3pt} \newline
\begin{tabular}{>{\centering\arraybackslash}m{32.5mm}m{32.5mm} }
Large $x \%$ $\alpha$ are $\beta$ and $\gamma$ & Large $x \%$ $\beta$ are $\alpha$ and $\gamma$ \\
Large $x \%$ $\gamma$ are $\alpha$ and $\beta$
\end{tabular} \\ \midrule

\end{tabular}
\caption{Re-formulation of graph concepts in terms of categorical propositions and extension to data triplets. It can be seen that graph concepts relying on partitioning $\mathcal{V}$ easily carry over to arbitrary data triplets.
} \vspace{-40pt}
\label{table_prop}
\end{table}

\begin{algorithm}[h!]
\scriptsize
  \caption{}\label{algo_prop}
  \begin{algorithmic}[1]
    \Procedure{get\_propositions}{$\mathcal{D}, x_{min}, y_{min}$}
    \State predicateList = [ ]
    \State propositionList = [ ]
     \For{$a_i \in A$ }
        \State predicateList.append ( get\_predicates( $\mathcal{D}_{(a_i, B, C)}, x_{min}, y_{min}$ ) )
      \EndFor
      \For{P $\in$ predicateList}
      	\State S $\gets$ merge\_subjects( P, predicateList ) \Comment{Combine all subjects ($a_i$) with predicate P}
        \State propositionList.append( (S, P) )
       \EndFor 
    \EndProcedure{ ( \textbf{return} propositionList ) }
    \Procedure{get\_predicates}{$\mathcal{D}_{(a_i, B, C)}, x_{min}, y_{min}$}
    \State regionList = $\mathcal{D}_{(a_i, B, C)}$ \Comment{Each triplet $(a_i, b_j, c_k)$ becomes a region ($b_j = \beta$, $c_k = \gamma$)}
    \State predicates = [ ]
    \While{ regionList } 
        \State grownRegionList $\gets$ grow\_regions( regionList ) \Comment{For each region, grow $\beta$ or $\gamma$ in dir. of max. density}
        \State regionList $\gets$ filter\_y( grownRegionList ) \Comment{Remove regions with density less than $y_{min}$}
        \State predicates.append( filter\_x( regionList ) ) \Comment{Store regions if at least $x_{min}$ fraction of triplets fall within} 
      \EndWhile\label{euclidendwhile}   
    \EndProcedure{ ( \textbf{return} predicates ) } \Comment{These predicates satisfy $x \geq x_{\min}$ $a_i$ are $y \geq y_{\min}$ P }
  \end{algorithmic}
\end{algorithm}

\end{document}